\documentclass[reprint, aps, superscriptaddress, prab]{revtex4-2}
\usepackage[utf8]{inputenc}
\usepackage{graphicx}
\usepackage{amsmath, amssymb}
\usepackage{xcolor}
\usepackage{upgreek}
\usepackage{bm}
\usepackage{siunitx}
\usepackage{hyperref}

\newcommand{\crit}{\mathrm{cr}}
\newcommand{\D}{\mathrm{d}}
\newcommand{\Eb}{\mathbf{E}}
\newcommand{\Bb}{\mathbf{B}}

\newcommand{\vb}{\mathbf{v}}

\newcommand{\ssb}{\mathbf{S}}
\newcommand{\Omegab}{\bm{\Omega}}

\newcommand{\nano}{\mathrm{nano}}
\newcommand{\rad}{\mathrm{rad}}

\begin{document}

    \title{Radiative depolarization of high-energy electron beams in wakefield accelerators}

    \author{Oliver Mathiak}
	\email{oliver.mathiak@hhu.de}
    \affiliation{Institut f\"{u}r Theoretische Physik I, Heinrich-Heine-Universit\"{a}t D\"{u}sseldorf, 40225 D\"{u}sseldorf, Germany}

	\author{Lars Reichwein}
	\email{l.reichwein@fz-juelich.de}
    \affiliation{Peter Gr\"{u}nberg Institut (PGI-6), Forschungszentrum J\"{u}lich, 52425 J\"{u}lich, Germany}
    \affiliation{State Key Laboratory of Ultra-intense Laser Science and Technology, Shanghai Institute of Optics and Fine Mechanics, Chinese Academy of Sciences, Shanghai 201800, People’s Republic of China}
	\affiliation{Institut f\"{u}r Theoretische Physik I, Heinrich-Heine-Universit\"{a}t D\"{u}sseldorf, 40225 D\"{u}sseldorf, Germany}

    \author{Alexander Pukhov}
	\affiliation{Institut f\"{u}r Theoretische Physik I, Heinrich-Heine-Universit\"{a}t D\"{u}sseldorf, 40225 D\"{u}sseldorf, Germany}
	
	 \author{Liangliang Ji}
	\affiliation{State Key Laboratory of Ultra-intense Laser Science and Technology, Shanghai Institute of Optics and Fine Mechanics, Chinese Academy of Sciences, Shanghai 201800, People’s Republic of China}

    \author{Markus B\"{u}scher}
	\affiliation{Peter Gr\"{u}nberg Institut (PGI-6), Forschungszentrum J\"{u}lich, 52425 J\"{u}lich, Germany}
	\affiliation{Institut f\"{u}r Laser- und Plasmaphysik, Heinrich-Heine-Universit\"{a}t D\"{u}sseldorf, 40225 D\"{u}sseldorf, Germany}

	\date{\today}
	
	\begin{abstract}
    The preservation of witness beam polarization in wakefield accelerators will be crucial for future collider applications. While extensive theoretical studies on the injection and initial acceleration of polarized electrons exist, a study concerning higher-energy regimes has been neglected thus far. Besides the spin precession usually considered in wakefield-related research, radiative effects could become increasingly relevant at higher energies as the witness electrons perform betatron oscillations during which they will emit photons. In the present study, we use particle-in-cell simulations extended with Monte-Carlo routines to study the influence of radiative spin-flips on beam polarization. We find that at high energies, the importance of radiative effects on beam polarization mainly comes down to the alignment of the witness beam with respect to the wakefield.
	\end{abstract}

\maketitle

\section{Introduction}
In the search for physics beyond the Standard Model, ever larger particle energies are required. Although conventional rf-based accelerators are well-established, their further development is limited by their size. In contrast, wakefield acceleration schemes can deliver much higher accelerating gradients, allowing for a significant compactification of the accelerator facility. In lieu of this, a new design initiative has been formed in the pursuit of realizing a 10 TeV particle center-of-mass collider \cite{Gessner2025_10TeV}. The experimental realization of those energies are still well out of reach, as highest energy achieved to date is on the level of $10$ GeV \cite{Picksley2024}.

In addition to high energies, many of the fundamental particle physics effects are sensitive to particle spin, driving a demand for high-quality polarized particle beams. Such applications include deep-inelastic scattering for probing the proton's nuclear structure \cite{Glashausser1979} or potentially for physics beyond the Standard Model for axion-like particle production \cite{Chen2025}. 

While such polarized beams are readily available for rf-based accelerators, producing them via wakefield acceleration remains an open problem. The main advantage of plasma-based polarized sources would be that these are not relying on additional infrastructure like the necessary photoguns which are not commonly accessible to many wakefield accelerator laboratories. Research on such plasma-based sources has only more recently gained traction. An overview of the progress on such sources is given in Ref. \cite{Reichwein2025}.

In addition to external injection of pre-polarized particle beams from rf-accelerators, two main approaches to generate polarized electron beams in a plasma have emerged. The first option is to pre-polarize a hydrogen halide gas using photo dissociation, and subsequently injecting and accelerating these electrons. This has been studied extensively in theoretical investigations, using different injection mechanisms including self-injection \cite{yin_generation_2024}, density down-ramp injection \cite{Wu2019_lwfa, Wu2019_pwfa}, colliding-pulse injection \cite{Bohlen2023} and pinching injection \cite{Reichwein2026}, yielding polarizations of up to $\sim 80 \%$ depending on the mechanism. As of now, for polarized electrons from plasma-based sources no experimental results have been produced, only for nuclear polarized helium-3 \cite{Zheng2024}. This is mainly due to restrictive target parameters concerning both density and interaction volume \cite{Sofikitis2025}. However, projects on experimental verification are still on-going \cite{Stehr2025}.
The second utilizes the spin-dependent ionization of atomic orbitals \cite{Nie2022}. While this method has the advantage of being a single-step process and does not require any additional target preparation, its realization is complicated as specific orbitals like $4f^{14}$ of ytterbium need to be ionized to yield $\sim 56 \%$.

Most of the current studies consider only classical spin precession, as described by the T-BMT equation \cite{Thomas1926, Bargmann1959}. However, when approaching higher energies, like the ones envisioned for the 10 TeV pCM collider, non-classical radiative effects become relevant. Most prominently, the non-linear inverse Compton effect will lead to significant radiation which may change the particles' spin. The strength of those strong-field QED effects are governed by the dimensionless quantum non-linearity parameter
\begin{align}
    \chi &= \frac{e \hbar}{m_e^3 c^4} \sqrt{-(F_{\mu\nu} p^\nu)^2} \; .
\end{align}
In the low-$\chi$ limit, this radiative spin change is called the Sokolov-Ternov effect and is commonly utilized in storage rings to achieve high degrees of polarization. In contrast, in wakefield accelerators, due to their different geometry and larger accelerating fields the same effect may lead to significant depolarization.
While Vieira \textit{et al.} showed in Ref. \cite{Vieira2011} that spin precession becomes negligible for high-energy electron beams, the same is not true for radiative polarization effects. In fact, the rate of photon emission and subsequent radiative spin change is expected to increase significantly with the particle energy, as $\chi$ scales directly with the energy. 
In the barely-quantum regime, where $\chi \ll 1$, the rate of photon emission scales with $\chi ^2$, so that even for moderate $\chi$ photon emission can substantially impact the polarization of the witness beam.
Moreover, while the Sokolov-Ternov effect in storage rings, where $\chi \ll 1$, predicts an equilibrium polarization of $8 / (5\sqrt{3}) \approx 92.4 \%$, this value depends on $\chi$ such that the influence of radiative effects could differ substantially in regimes relevant to research on quantum electrodynamics \cite{Baier1972}.

In this paper, we investigate the contribution of such radiative spin-flips in high-energy wakefield accelerators using particle-in-cell (PIC) simulations. A pre-polarized electron beam is placed in a wakefield structure, and the influence of both spin precession and radiative spin flips on the beam polarization are tracked. In the following, in section \ref{sec:theory}, we first detail the relevant spin-dependent effects in our regime. In section \ref{sec:pic}, we detail our simulation setup, as well as the results of various parameter scans. The results are discussed in section \ref{sec:discussion}.

\section{Spin-dependent effects in strong fields} \label{sec:theory}

In the classical limit, where $\chi \ll 1$, the spin dynamics are described by the T-BMT equation \cite{Thomas1926, Bargmann1959},
    \begin{align}
        \frac{\D \ssb}{\D t} = - \Omegab \times \ssb \; ,
    \end{align}
    with the precession frequency
    \begin{align}
        \Omegab = \frac{e}{mc} \left[ \Omega_B \Bb - \Omega_V \left( \frac{\vb}{c} \cdot \Bb \right) \frac{\vb}{c} - \Omega_E \frac{\vb}{c} \times \Eb \right] \; , \label{eq:prec}
    \end{align}
where $e$ denotes the elementary charge, $m$ the electron mass, $c$ the vacuum speed of light, $\vb$ the particle velocity and $\Eb, \Bb$ the electromagnetic field. The three prefactors are defined as
    \begin{align}
        \Omega_B = a + \frac{1}{\gamma} \; , && \Omega_V = \frac{a \gamma}{\gamma + 1} \; , && \Omega_E = a + \frac{1}{\gamma + 1} \; ,
    \end{align}
with the anomalous magnetic moment $a \approx 10^{-3}$ for electrons and the Lorentz factor $\gamma$. This description of the spin dynamics is used in most wakefield-related studies e.g. \cite{Wu2019_lwfa, Sofikitis2025}, where fields and particle energies are sufficiently low.

\begin{figure}[h]
    \centering
    \includegraphics[width=0.5\textwidth]{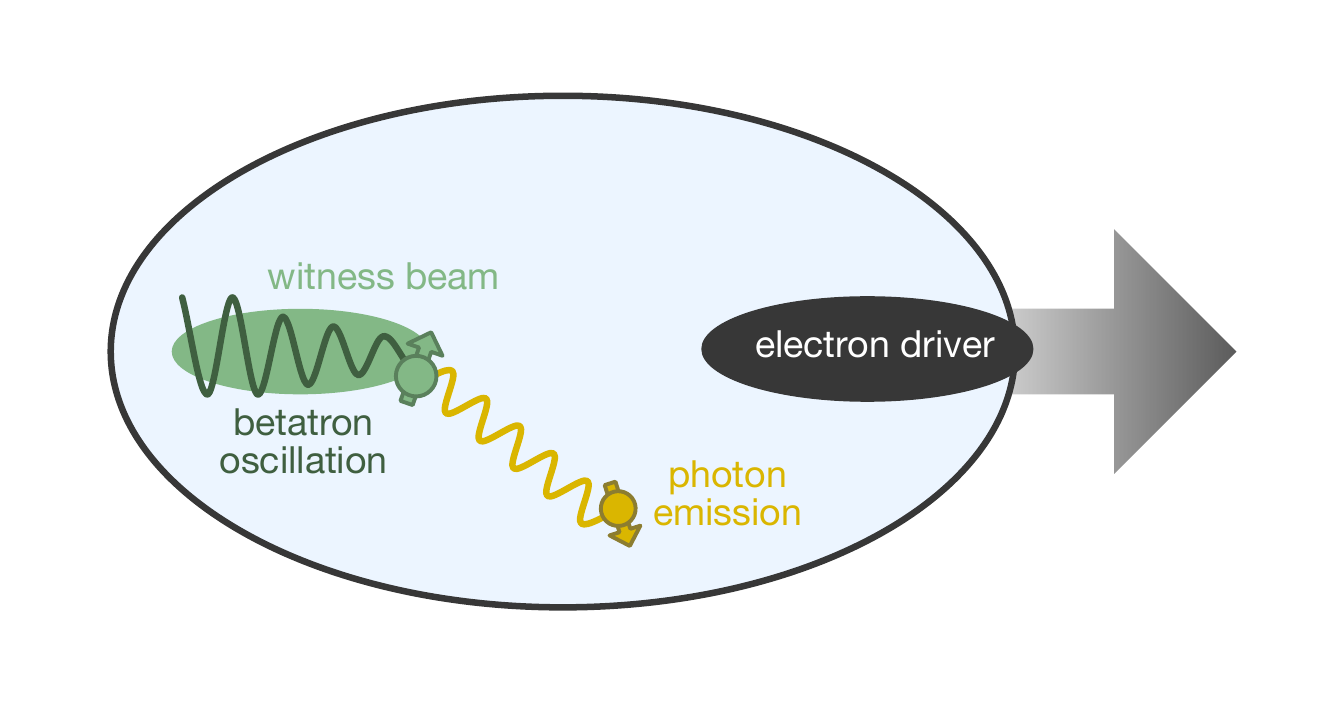}
    \caption{Schematic depiction of the electron driver (gray) generating a blowout in the plasma. The pre-polarized witness electrons (green) perform betatron oscillations during which they can emit photons (yellow), potentially leading to spin flips.}
    \label{fig:overview}
\end{figure}

However, as the particle energy increases and $\chi$ approaches unity, strong-field QED effects become relevant. In the strong fields of the blowout, the particles will undergo betatron oscillations during which they will emit photons (cf. Fig. \ref{fig:overview}).
While most descriptions of the non-linear inverse Compton effect do not consider the particle's spin or the polarization of the emitted photon, in general the rate of emission is dependent on both. Thus, when the field strength in the rest frame of the particle approaches the critical Sauter-Schwinger field $ B_{\crit} = m^2c^3/ (e\hbar)$, the photon emission becomes strong enough to make Compton-induced spin change relevant. 

When emitting a photon, the emitting electron has a chance to flip its spin state along some defined quantization axis. The fully spin- and polarization-resolved differential emission rate of Compton photons \cite{Song2022} can be written as
\begin{align}\label{eq:dif_rate}
    \frac{\D^{2}W_{\rad}}{\D u \D t}&=\frac{C_{\rad}}{4}(w_{\rad}+\langle\mathbf{g}_{\rad}, \mathbf{S}_{i}\rangle+\langle \mathbf{p}_{f}, \mathbf{S}_{f} \rangle+ \langle \mathbf{p}_{\xi},\mathbf{\xi} \rangle),
\end{align}
Here, $C_{\rad} = \alpha m_e^2 c^4/(\sqrt{3}\pi\hbar\varepsilon_e)$, with the fine-structure constant $\alpha = e^2/\hbar c\approx 1/137$, the electron rest mass $m_e$, the speed of light in vacuum $c$, the reduced Planck constant $\hbar$ and the electron energy $\varepsilon_e$. The rate is given in terms of the energy ratio $u=\hbar\omega /\varepsilon_e$ transferred from the emitting electron to the photon. The term
\begin{align}
    {w_{\rad}=\frac{u^{2}-2u+2}{1-u}K_{2/3}(y)-\mathrm{Int}K_{1/3}(y),}
\end{align}
describes the spin and polarization independent contribution to the photon emission rate. In fact, the commonly used spin-and polarization averaged rate is $\frac{\D^2\overline{W}_{\rad}}{\D u \D t} = C_{\rad}w_{\rad}$. 
Here, $K_{\nu}$ is the modified Bessel function of the second kind, $\mathrm{Int} K_{1/3}(y) = \int_y^\infty K_{1/3}(x) \;  \D x$ and $y = 2u/[3(1-u)\chi_3]$.
The second term in \eqref{eq:dif_rate} is a correction due to the initial spin $\mathbf{S}_i$ of the electron before photon emission and is given by
\begin{align}
    \mathbf{g}_{\rad}=-uK_{1/3}(y_{1})\mathbf e_{2}.
\end{align}
The spin corrections are given in terms of the orthonormal basis $(\mathbf{e}_1, \mathbf{e}_2, \mathbf{e}_v)$, which is defined as
\begin{align}
    \mathbf{e}_v = \frac{\mathbf{p}_e}{|\mathbf{p}_e|} \; , && \mathbf{e}_1 = \mathbf{F}_L - \langle \mathbf{F}_L, \mathbf{e}_v \rangle \mathbf{e}_v \; ,&& \mathbf{e}_2 = \mathbf{e}_v \times \mathbf{e}_1 \; .
\end{align}
Here, $\mathbf{e}_v$ is the forward direction of the emitting electron, $\mathbf{e}_1$ the direction of the transverse part of the Lorentz force and the corresponding normal vector $\mathbf{e}_2$.
In addition to the initial spin $\mathbf{S}_i$, the photon emission rate also depends on the final spin state $\mathbf{S}_f$ of the electron. This contribution is described by the third term in \eqref{eq:dif_rate} with
\begin{align}
    \mathbf{p}_{f}&=-\frac{u}{1-u}K_{1/3}(y)\mathbf{e}_{2}+\left[2K_{2/3}(y)-\mathrm{Int}K_{1/3}(y)\right]\mathbf{S}_i \notag \\
    &\quad+\frac{u^2}{1-u}\left[K_{2/3}(y)-\mathrm{Int}K_{1/3}(y)\right]\langle \mathbf{S}_i, \mathbf{e}_v\rangle \mathbf{e}_v \; .
\end{align}
This dependence on the final spin leads to a change of spin with each photon emission. When considering mixed-state representation of the particles spin, where $|\mathbf{S}| \le 1$, the state after photon emission is given by
\begin{align}
    \mathbf{S}_f = \mathbf{p}_f  / [C_{\rad}(w_{\rad} + \langle \mathbf{g}_{\rad}, \mathbf{S}_i \rangle)] \; .
\end{align}

In addition to the change of spin due to the photon emission, due to the dependence of emission on the electron spin, we get non-radiative spin polarization when no photon is emitted,
\begin{align}
    \mathbf{S}_f^\mathrm{NR} = \frac{\mathbf{S}_i \left(1- C_{\rad}\Delta t\int_0^1w_{\rad} \D u\right) - C_{\rad} \Delta t \int_0^1 \mathbf{g}_{\rad} \D u}{1-C_{\rad} \Delta t \int_0^1 \left(w_{\rad} + \langle \mathbf{g}_{\rad}, \mathbf{S}_i \rangle\right) \D u } \; .
\end{align}
    
The final term in \eqref{eq:dif_rate} couples the emission rate with the polarization of the emitted photon. Here, $\bm{\xi} = (\xi_1, \xi_2, \xi_3)$ is the Stokes vector of the photon, which consists of the Stokes parameters. The scaled quantization axis is given by
\begin{align}
        \mathbf{p}_\xi &= \frac{u}{1-u}  K_{1/3}(y) \langle \mathbf{S}_i,\mathbf{e}_1\rangle \mathbf{e_1} \notag \\ 
        &+ \left[ \frac{2u-u^2}{1-u}K_{2/3}(y) - u \mathrm{Int}K_{1/3}(y) \right] \langle \mathbf{S}_i, \mathbf{e}_v \rangle \mathbf{e}_2 \notag \\
        &+ \left[K_{2/3}(y) - \frac{u}{1-u}K_{1/3}(y) \langle \mathbf{S}_i, \mathbf{e}_2 \rangle
        \right] \mathbf{e}_v \; .
\end{align}
In the following, we will study the influence of both precession and radiative processes and the total polarization of a witness beam.

\section{Particle-in-cell simulations} \label{sec:pic}
For our simulations, we use the particle-in-cell (PIC) code \textsc{vlpl} \cite{Pukhov1999, Pukhov2016}. It has been extended to incorporate both spin precession according to the T-BMT equations as well as radiative processes. 
The simulation domain has a size of $125 \si{\micro\meter} \times 75 \si{\micro\meter}$, with a resolution of $h_x = 0.125 \si{\micro\meter}$, $h_y = 0.25 \si{\micro\meter}$ and $\Delta t = h_x / c$, in accordance with the rhombi-in-plane Maxwell solver \cite{Pukhov2020}.

The setup is as follows: the driver is a double-Gaussian electron beam with length $\sigma_x = 5 \si{\micro\meter}$, radius $\sigma_r = 1 \si{\micro\meter}$ and charge $2.25 \si{\nano\coulomb}$. 
It has an energy of 500 GeV and propagates in $x$-direction through a background plasma, generating a wakefield.
The background plasma is modeled as a homogeneous electron gas with $n_b = 2.2 \times 10^{18} \si{\per\centi\meter\cubed}$ and an initial linear density up-ramp of $\sim 400 \si{\micro\meter}$.
The accelerated witness beam is placed in the back part of the blowout ($\sim 66 \si{\micro\meter}$ behind the driver) and initially fully polarized.
Its parameters like initial energy and transverse size will be changed according to the different parameter scans detailed in the following.
All species are simulated with 16 particles per cell.

\subsection{Precession vs. radiation}

We investigate the influence of spin precession and radiative effects, by running different simulations considering either effect alone and both activated at the same time. Note, that the effects are not expected to be additive, since a change in spin due to precession will lead to a different initial spin for the eventual photon emission event. Likewise, if a spin-flip occurs, the electron will now exhibit a spin value that will precess differently from its original orientation in a given external field.

\begin{figure}[h]
    \centering
    \includegraphics[width=0.5\textwidth]{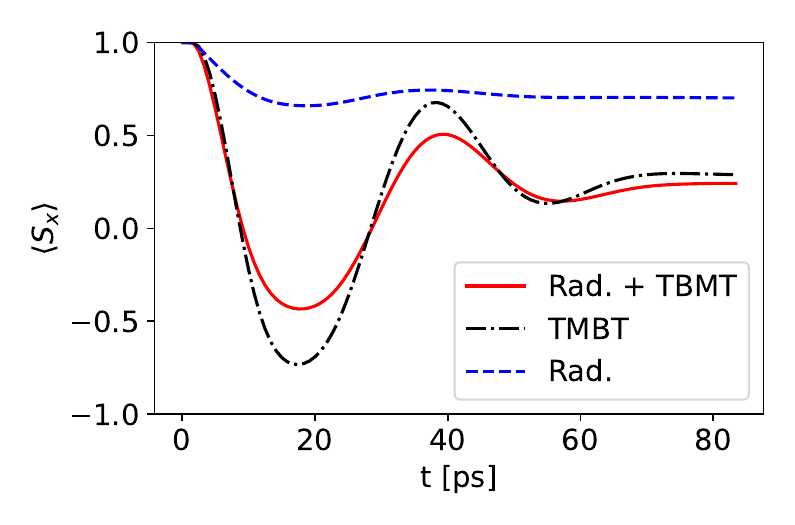}
    \includegraphics[width=0.5\textwidth]{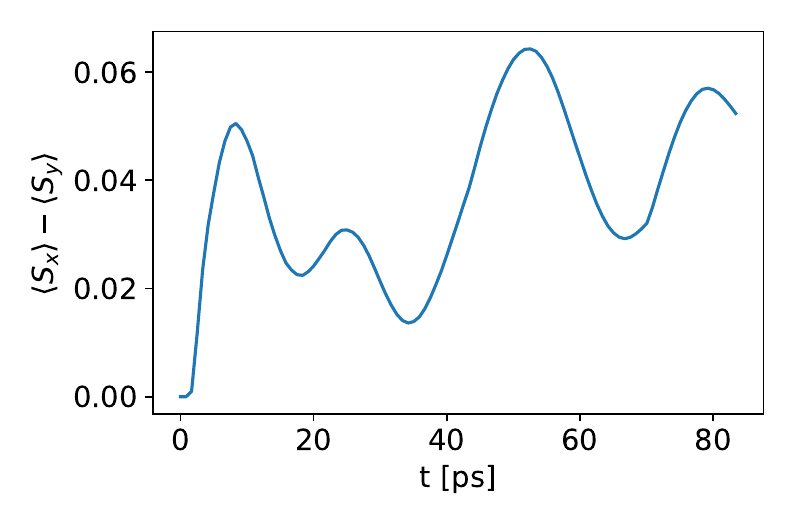}
        \caption{Top: Comparison of spin evolution for a full simulation (red line), only T-BMT precession (black dashed-dotted), and only radiative effects (blue dashed) for initial longitudinal polarization. Bottom: Difference between simulations with initial longitudinal and transverse polarization. Here, both T-BMT and radiative processes are considered.}
    \label{fig:cmp_effects}
\end{figure}

The witness beam has an initial energy of 1 TeV and a radius of $\sigma_x = \sigma_r = 2.5 \si{\micro\meter}$.
We consider both initial longitudinal polarization, $\langle s_x \rangle = 1$, and transverse polarization $\langle s_y \rangle = 1$. To make the influence of radiative effects more apparent, we displace the witness beam with respect to the optical axis by $\Delta y = 10 \si{\micro\meter}$, leading to stronger transverse fields and subsequent $\chi$ parameter.
The spin evolution is shown in Fig. \ref{fig:cmp_effects}.

When considering only T-BMT precession, we see the expected oscillations in the spin polarization. The amplitude of these oscillations reduces over time and will converge to a fixed value, the final polarization.
Only considering radiative spin change, we observe an initial decrease in polarization to approx. 0.6 in 20 ps, regardless of initial polarization direction. Afterwards, the polarization value remains largely constant. Considering the combined effects, we observe that the contribution from the T-BMT precession is dominant. The radiative contribution dampens the amplitude of the oscillation. The difference between initially longitudinally and transversely polarized witness beams is only minor and is on the order of multiple percent. A comparable discrepancy between longitudinal and transverse initial polarization is expected for only the T-BMT precession in the low-$\chi$ regime as well.

\subsection{Beam energy}

\begin{figure}[h]
    \centering
    \includegraphics[width=0.5\textwidth]{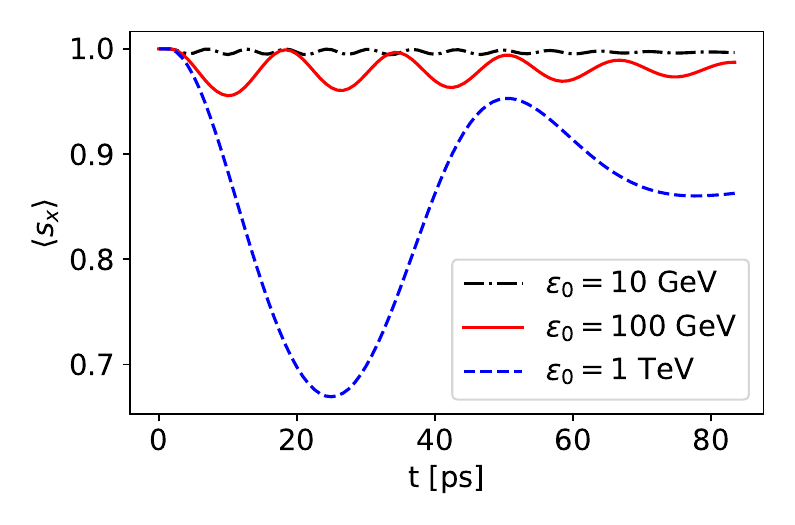}
    \caption{Influence of initial witness energy on the beam polarization over time. Note that the witness is injected into the same phase of the wakefield for all cases, which can affect the initial photon emission.}
    \label{fig:energy}
\end{figure}

To investigate the dependence on the witness energy, we vary the initial energy $\varepsilon_0$, for a beam with $\sigma_r = \sigma_x =  2.5 \si{\micro\meter}$, and charge $1.3 \si{\nano\coulomb}$. Here, we consider no transverse offset, i.e. $\Delta y = 0$. 
The results are shown in Fig. \ref{fig:energy}. We see that the highest-energy beam exhibits the most significant depolarization, whereas the lower-energy beam remains almost fully polarized. 
The oscillation frequency is inversely proportional to the energy.
One crucial detail here is how well the witness injection is matched to the driving wakefield. Depending on the energy of the witness beam, and its dimensions as well as position with respect to the wakefield, it could radiate a significant part of its energy immediately. In turn, radiative depolarization could outweigh any precession contributions for strongly mismatched beams.

\subsection{Beam radius and transverse offset}

The focusing fields of the blowout regime are linear in $r$. Consequently, $\chi$ grows identically, leading to stronger radiation and radiative polarization. The same is true for the precession of the particles' spin. Therefore, we expect the outer particles of the beam to be subject to stronger spin changes. As such, the depolarization rate of the beam is expected to depend on both the radius of the witness beam as well as a potential offset from the propagation axis.
To investigate, we conduct simulations with a witness beam of the same shape as above with $\varepsilon_{0} \sim 100$ GeV with its transverse size being varied between different simulation runs.

As expected, the beam with the largest transverse diameter exhibits the strongest depolarization in the simulations. 
If the beam has a transverse offset $\Delta r \neq 0$ with respect to the optical axis, the polarization decreases even further. As an example, we repeat the simulation with $\sigma_y = 2.5 \si{\micro\meter}$, with a transverse displacement of $\Delta y = 10 \si{\micro\meter}$.

\begin{figure}[h]
    \centering
    \includegraphics[width=0.5\textwidth]{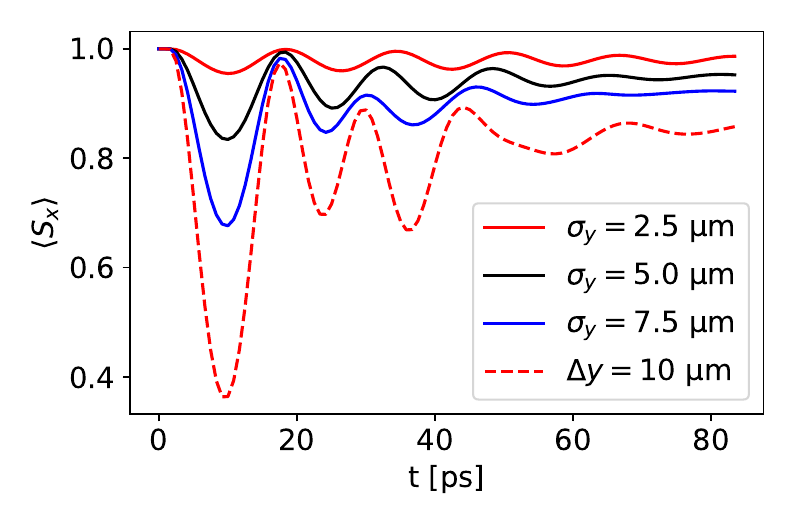}
    \caption{Influence of the beam radius on the polarization evolution over time. The red dashed line shows the behavior of a beam with $\sigma_y = 2.5 \si{\micro\meter}$, but with a transverse offset of $10 \si{\micro\meter}$.}
    \label{fig:radius}
\end{figure}

\section{Discussion} \label{sec:discussion}

\begin{figure*}[th]
    \includegraphics[width=\textwidth]{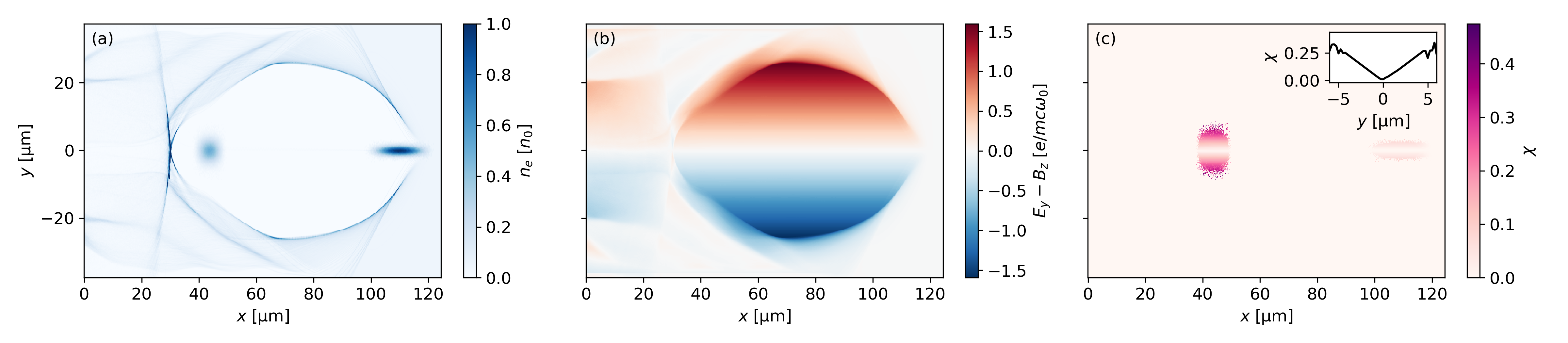}
    \caption{\label{fig:2d_chi} An exemplary simulation showing the (a) density distribution, including both driver ($x \sim 110 \si{\micro\meter}$) and witness beam ($\sim 45 \si{\micro\meter}$). Subplot (b) shows the radial dependence of the focusing force, and (c) the dependence of $\chi$. The inset underlines the linear increase of $\chi$ with the radial position.}
\end{figure*}

Our simulations underline the need to properly align the witness beam with the blowout to retain polarization. This is a direct result of the structure of the wake's fields and the dependence of both effects on the field amplitude.

In the simplest approximation \cite{Kostyukov2004}, the electromagnetic fields inside the blowout are
\begin{align}
    E_x = \frac{\xi}{2} \; , && E_y = - B_z = \frac{y}{4} \; , \notag \\
    B_x = 0 \; , && E_z = B_y = \frac{z}{4} \; .
\end{align}
Here, $\xi = x - t$ is the co-moving coordinate. The fact that the focusing fields are linear holds also according to more general theoretical models; the most significant changes in the field structure compared to the simplified model are expected in the longitudinal electric field due towards the bubble back \cite{Reichwein2020}.
In Fig. \ref{fig:2d_chi}, we show the placement of the witness beam in the wakefield, as well as the distribution of the focusing force and the $\chi$-value across the blowout.

For the precession frequency \eqref{eq:prec} of the T-BMT equations, we see that the terms will scale linearly with the radial offset $r$ of the beam since the terms are proportional to the fields $\Eb, \Bb$.

Considering, the radiative processes, for small $\chi \ll 1$ and constant energy, we may approximate the rate of spin change by 
\begin{align}
    \frac{\D S_x}{dt} = - \frac{3\chi^2}{4\tau_\mathrm{ph}} \left(S_x + \frac{8}{5\sqrt{3}} \right)
\end{align}
with the characteristic time for photon emission $\tau_\mathrm{ph}  = 2/(5\pi C_{\rad}\chi)$ \cite{Artemenko2023}. Thus, the rate of spin change due to photon emission is $\propto \chi^3$ and accordingly to $r^3$. As such, the polarization of the beam is highly sensitive to the beam radius and alignment along the propagation axis.

Beam polarization is far from the only concern for TeV-class wakefield accelerators (see, e.g., Qian \textit{et al.} \cite{Qian2026}). Nonetheless, its preservation will be crucial for various physics applications. Besides reducing the offset, the polarization quality of such beams could be increased by utilizing spin filters as proposed by Wu \textit{et al.} \cite{Wu2020}.

\section{Conclusions}

We have studied the influence of spin precession according to the T-BMT equations as well as radiative spin-flips for high-energy, polarized electron beams in wakefields. The particle-in-cell simulations show that proper alignment of the injected beams with respect to the accelerating wakefield will be necessary to mitigate depolarizing contributions from both spin precession and radiative effects.
A next step will be to experimentally verify that beam polarization is preserved upon injection of a pre-polarized beam into a wakefield. Such a result, even for lower energies is still outstanding, but could be realized via both conventional polarized sources or further development of plasma-based schemes such as \cite{Nie2022, Sofikitis2025}.

\begin{acknowledgments}
O.M. and L.R. contributed equally to this manuscript.
    The authors gratefully acknowledge the Gauss Centre for Supercomputing e.V. \cite{GCS} for funding this project (spaf) by providing computing time through the John von Neumann Institute for Computing (NIC) on the GCS Supercomputer JUWELS at J\"ulich Supercomputing Centre (JSC).
\end{acknowledgments}

\bibliography{bib_radiative.bib}

\end{document}